\documentclass[preprint,prd,nofootinbib,tightenlines,amsmath]{revtex4}

\usepackage{epsfig}
\usepackage{graphics,color}      % usual driver
\usepackage{verbatim}
\usepackage{amsmath}    % need for subequations
\catcode`\@=11
\def\sla@#1#2#3#4#5{{%
 \setbox\z@\hbox{$\m@th#4#5$}%
 \setbox\tw@\hbox{$\m@th#4#1$}%
 \dimen4\wd\ifdim\wd\z@<\wd\tw@\tw@\else\z@\fi
 \dimen@\ht\tw@
 \advance\dimen@-\dp\tw@ \advance\dimen@-\ht\z@
 \advance\dimen@\dp\z@
 \divide\dimen@\tw@ \advance\dimen@-#3\ht\tw@
 \advance\dimen@-#3\dp\tw@ \dimen@ii#2\wd\z@
 \raise-\dimen@\hbox to\dimen4{%
 \hss\kern\dimen@ii\box\tw@\kern-\dimen@ii\hss}%
 \llap{\hbox to\dimen4{\hss\box\z@\hss}}}}

\def\slashed#1{%
 \expandafter\ifx\csname sla@\string#1\endcsname\relax
{\mathpalette{\sla@/00}{#1}}
\fi}
\def\declareslashed#1#2#3#4#5{%
 \expandafter\def\csname sla@\string#5\endcsname{%
#1{\mathpalette{\sla@{#2}{#3}{#4}}{#5}}}}
 \catcode`\@=12
\declareslashed{}{/}{.08}{0}{D}
 \declareslashed{}{/}{.1}{0}{A}
 \declareslashed{}{/}{0}{-.05}{k}
 \declareslashed{}{/}{.1}{0}{\partial}
 \declareslashed{}{\not}{-.6}{0}{f}

\def\lsim{\mathrel {\vcenter {\baselineskip 0pt \kern 0pt
    \hbox{$<$} \kern 0pt \hbox{$\sim$} }}}
\def\gsim{\mathrel {\vcenter {\baselineskip 0pt \kern 0pt
    \hbox{$>$} \kern 0pt \hbox{$\sim$} }}}

\begin{document}

\baselineskip=15pt

\preprint{BNL-HET-08/7}

\hspace*{\fill} $\hphantom{-}$

\title{Towards establishing the spin of warped gravitons }

\author{Oleg Antipin}
\email{oaanti02@iastate.edu}
\affiliation{Department of Physics and Astronomy, Iowa State University, Ames, IA 50011, USA}

\author{Amarjit Soni}
\email{soni@bnl.gov}
\affiliation{Brookhaven National Laboratory, Upton, NY 11973, USA\\}

\date{\today}

\begin{abstract}     
                  
We study the possibility of experimental verification of the spin=2 nature 
of the Kaluza-Klein (KK) graviton which is predicted to exist in the 
extra-dimensional Randal-Sundrum (RS) warped models. The couplings of these 
gravitons to the particles located on or near the TeV brane is the strongest 
as the overlap integral of their profiles in the extra-dimension is large. 
Among them are unphysical Higgses ($W^{\pm}_L$ and $Z_L$) and KK excitations 
of the Standard Model (SM) gauge bosons. We consider the possibility to 
confirm the spin-2 nature of the first KK mode of the warped graviton ($G_1$) 
based on the angular distribution of the Z bozon in the graviton rest frame 
in the gg$\to G_1 \to W^{KK} (Z^{KK}) W (Z)\to WWZ$, gg$\to G_1\to ZZ$ and 
gg$\to G_1 \to Z^{KK} Z\to ZZH$ decay channels. Using Wigner D-matrix properties, 
we derive the relationship between the graviton spin, signal angular 
distribution peak value, and other theoretically calculable quantities. 
We then study the LHC signals for these decay modes and find that with 
1000 fb$^{-1}$ of data, spin of the RS graviton up to $\sim$ 2 TeV may be 
confirmed in the $pp \to W^{KK} (Z^{KK}) W (Z) \to WWZ \to$ 3 leptons + 
jet + $\slashed{E}_T$ and $pp \to  ZZ \to$ 4 leptons decay modes.

\end{abstract}

\pacs{PACS numbers: }

\maketitle

\section{Introduction}

With the upcoming start of the CERN LHC, our quest for the physics beyond SM is 
likely to yield positive results. On the theoretical side two of the most important 
questions to be answered are the Planck-weak hierarchy problem and the flavor puzzle 
of the SM. The Randall-Sundrum model with a warped extra dimension \cite{Randall:1999ee} 
is just about the only theoretical framework which simultaneously addresses both these 
questions making it a very compelling model of new physics. Perhaps the most distinctive 
feature of this scenario is the existence of KK gravitons with masses and couplings at 
the TeV scale which therefore should appear in experiment as widely separated resonances
\cite{Davoudiasl:2000wi}.

The original RS model as well as all of its extensions are based on a slice of Ad$S_5$ 
space. At the endpoints of this five-dimensional space ($\phi=0,\pi$), two branes are placed which are 
usually labeled as an ultraviolet (UV) Planck brane and an IR (TeV) brane; and the large 
hierarchy of scales is solved by a geometrical exponential factor. Postulating modest-sized 
$5^{th}$ dimension with radius R and curvature k the TeV/Planck $\sim e^{-k\pi R}$  ratio 
of scales can be numerically obtained by setting $kR$ $\approx$11. In the original RS model 
all SM fields were localized on the TeV brane. The only new particles in this model were KK 
gravitons with no SM gauge quantum numbers. Later, in addition to the KK gravitons, 
a bulk scalar field with a $\phi$-dependent vacuum expectation value (VEV) 
was shown to generate a potential to stabilize the $R$ modulus \cite{Goldberger:1999uk}. 
However, this model leaves higher-dimensional 
operators in the 5D effective field theory suppressed only by TeV scale which, in turn, 
generates unacceptably large contributions to flavor changing neutral current (FCNC) and 
observables related to the SM electroweak precision tests (EWPT). 
A natural way to avoid this problem, proposed by 
\cite{Davoudiasl:1999tf,Pomarol:1999ad,Grossman:1999ra,Huber:2000ie,Gherghetta:2000qt}, 
is to allow SM fields to propagate in the extra dimension. In this scenario there are KK 
excitations of SM gauge and fermion fields in addition to those of the graviton. These states 
have masses in the TeV range and are localized near the TeV brane. The SM particles are the 
zero-modes of the 5D fields, and the profile of a SM fermion in the extra dimension depends 
on its 5D mass. By localizing light fermions near the Planck brane and heavier ones near the 
TeV brane, the contributions to the FCNC and EWPT are suppressed by scales $\gg$ TeV. As a 
consequence, the KK graviton whose profile is peaked at the TeV brane will couple mostly 
to the top quark, Higgs (or, by equivalence theorem, to the longitudinal $W$ and $Z$ bosons)
~\cite{Fitzpatrick:2007qr,Agashe:2007zd,Davoudiasl:1999jd}, and KK excitations of the SM fields.

Thus, the promising channels to observe RS gravitons are those where produced gravitons are 
decaying to fields localized near TeV brane. The search for the KK gravitons using its decays 
to the top quarks was performed in \cite{Fitzpatrick:2007qr}. Signals from graviton decay to 
$W_L$ pair, which subsequently decay into pure leptonic or semileptonic final states, were 
considered in \cite{Antipin:2007pi}. The 4-lepton signal through the decay to a pair of $Z_L$'s 
was studied in \cite{Agashe:2007zd}. Reconstruction possibility of the $Z$'s via their leptonic 
decays makes this a uniquely clean mode. 

In this paper we would like to address the issue of confirmation of the spin-2 nature of RS 
gravitons (for the most recent survey of methods of measuring the spin of new physics particles 
at the LHC, see \cite{Wang:2008sw}). The conventional way to measure the spin of a new particle 
involves reconstruction of its rest frame using its decay products and studying the angular 
distributions about the polarization axis. Along these lines, in \cite{Fitzpatrick:2007qr} 
generic sample of a 100 $t\bar{t}$ events was produced for a spin-0, spin-1 and spin-2 resonances 
in an attempt to distinguish spin of the resonance based on the angular dependence of the 
cross-section. We, instead, 
concentrate on the gg$\to G_1 \to W^{KK} (Z^{KK}) W_L (Z_L) \to$ ZZH,WWZ and gg$\to G_1 \to Z_L Z_L$ 
channels where, in addition to the $Z_L Z_L$  channel considered in the literature before, we have 
one KK partner of the W or Z and one longitudinally polarized W or Z in the final state. Due to 
$W^{KK}$ and $Z^{KK}$ presence in these new channels, invariant mass of their decay products should 
show resonant behavior. Reconstruction of these intermediate KK gauge bosons will be important to 
reveal the internal structure of the RS model. We will see that Z boson signal angular distribution 
in the graviton rest frame for all these modes peaks at 90 degrees to collision axis. Performing 
angular analysis using  Wigner D-matrix, we will derive the relationship between the graviton spin, 
angular distribution peak value, and other theoretically calculable quantities. As our method only 
requires to measure this peak value, where most of the signal events will be concentrated, we may 
optimistically achieve this goal with a relatively low sample of $O$(10) events.

\section{Model}
We closely follow the model discussed in \cite{Agashe:2007zd} and briefly 
review it here. As discussed above, we allow SM fields to propagate in the extra-dimension and 
distribute fermions along it to generate observed mass spectrum without introducing additional 
hierarchies in the fundamental 5D theory. SM particles are identified with zero-modes of 5D fields, 
and the profile of the fermion in the extra dimension depends on its 5D mass. As was shown before 
\cite{Grossman:1999ra,Huber:2000ie,Gherghetta:2000qt}, all fermion 5D masses are $O$(1) parameters 
with the biggest one, among the SM quarks, being that of the top quark. To specify the model even 
further, the top quark is localized near the TeV brane and the right-handed isospin is gauged 
\cite{Agashe:2003zs}. We consider $t_R$ being on the TeV brane (see discussion of the other 
possibilities in \cite{Agashe:2007zd}, for example). At the end of the day, we are left with three 
parameters to be measured experimentally. We define them as $c\equiv k/M_{Pl}$ (the ratio of the 
AdS curvature $k$ to the Planck mass), $\mu\equiv ke^{-\pi kR}$ 
monitors gauge KK masses with the first few being (2.45, 5.57,
8.7...)$\times \mu$, and finally the parameter $\nu\equiv m/k$ which defines where the lightest 
fermion with bulk mass $m$ is localized. For the $t_R$ on the TeV brane, $\nu_{t_R}\approx0.5$; 
and parameters $c$ and $\mu$ will remain free in our analysis.

\subsection{Low energy constraints on model parameters}

Before proceeding further, let us briefly review constraints 
placed on the warped extra-dimension 
model with custodial isospin symmetry \cite{Agashe:2003zs}, 
which we adopt in this paper. In the resulting setup of \cite{Agashe:2003zs},
KK mass scale as low as $\sim$ 3 TeV is allowed by precision electroweak 
data. 
Regions of parameter space that successfully reproduce the fit
to electroweak precision observables with KK excitations
as light as $\sim$ 3 TeV were also studied in \cite{Carena:2006bn}.
Phenomenological consequences of the observed  
$B\bar{B}$-mixing were discussed in \cite{Chang:2006si}. 
In the model of \cite{Agashe:2003zs} tree level exchange of KK gluons gives 
the dominant contribution to the 
$B\bar{B}$-mixing. In \cite{Chang:2006si}, 
the CP-violating effects on the $B_d$ system were shown to 
provide  $M^{(1)}_{gluon}>$3.7 TeV  constraint at 68\% CL.

Phenomenological constraints from lepton-flavor-violations were
discussed in \cite{Agashe:2006iy, Chang:2007uz}. In
\cite{Agashe:2006iy}, ``anarchic'' Randall-Sundrum model of flavor was
studied, and the minimal allowed KK scale of $\sim$ 3 TeV was found to
be permitted for a few points in the natural RS parameter space; but
models with custodial isospin can relax these constraints. 
In
\cite{Chang:2007uz}, extensive analysis of $B \to K^* l^+ l^{'-}$
modes was performed, concluding that only the $B\to K^* ee$ decay have 
sizable new physics effects. With SM contributions being suppressed, 
current experimental bounds were translated into the lepton bulk mass 
parameters. For the first KK
gauge boson mass of 2-4 TeV, 10-20$ \%\ $ deviation from the SM results
were found. Top quark flavor violations
and B-factory signals were also studied 
in \cite{Agashe:2004ay,Burdman:2003nt,Agashe:2006wa}.  
Finally, enhanced contributions to $\Delta S=2$ processes 
generated by beyond the SM operators with $(V - A) \otimes (V + A)$ structure,
present in these frameworks, may impose additional constraints  \cite{Beall:1981ze}.  
%and within our framework they are $\sim$ $m_d m_s$ \cite{Agashe:2004ay}. 
Without further flavor structure these contributions were expected to place a 
lower bound on the KK gluon 
mass of $O$(8 TeV)~\cite{Bona:2007vi,Agashe:2007ki}. However, most recent studies of 
the flavor constraints on the new physics mass scale find that 
the KK gluon mass should generically be heavier than about 21 TeV \cite{Csaki:2008zd}.

Relentless attempts to lower KK-mass scale further still flourish on the market. 
On this road, a number of other models were proposed trying to improve the prospects to discover 
KK-particles at the LHC. 
One of them is a model presented in \cite{Moreau:2006np} with a 
somewhat surprising claim
that KK masses as low as 1 TeV are consistent with all
current experimental constraints. An interesting variant of the
warped extra dimension based on 5D minimal flavor violation was
recently proposed in \cite{Fitzpatrick:2007sa}. The model allows to
eliminate current RS flavor and CP problem ~\cite{Agashe:2004ay,Agashe:2004cp} with a KK scale as low as 2
TeV. Closing the list of examples, a volume-truncated version of the RS scenario called 
``Little Randall-Sundrum (LRS)" model 
was constructed in \cite{Davoudiasl:2008hx}. With the assumption of separate gauge 
and flavor dynamics, this setup allows to suppress 
a number of unwanted contributions to precision electroweak, 
$Z b\bar b$, and flavor observables, compared with the corresponding RS case.

Summarizing, we may say that KK gauge bosons with masses below 3 TeV 
(which would imply $m_G \gsim$ 4
TeV) would be difficult to have in current theoretical constructions. 
If this is the case, signals at the LHC, confirming the RS idea,
would be extremely difficult to find; and studies conducted in 
~\cite{Agashe:2007zd,Agashe:2007ki} and later in this paper support this
unfavored future.
However, in view of the above discussion, it also seems plausible
that these models are still being developed; and, therefore,
it is not inconceivable that explicit construction(s) will be found
which will allow KK masses lower than 3 TeV without conflict with 
electroweak precision experiments and/or with flavor physics.
This attitude was taken in \cite{open} and we in this paper 
will also adopt this point of view.

\subsection{Couplings of KK gravitons}

After these brief remarks we can write the couplings relevant to our discussions here. 
Since the graviton $h_{\mu\nu}$  couples to the energy-momentum tensor $T^{\mu\nu}$, 
coupling  of the nth level KK graviton to the qth and mth level gauge bosons has the 
generic form: 
\begin{equation}
L_G=\frac{C_{qmn}}{M_{Pl}}T^{\mu\nu(q,m)} h_{\mu\nu}^{(n)},
\end{equation} 
where the magnitude of the $C_{qmn}$ coupling constants depends on the overlap of 
the particle wavefunctions in the extra-dimension.

Analytic expressions for the coefficients $C_{qmn}$ with the flat zeroth mode 
gauge boson profile may be found in \cite{Davoudiasl:2000wi} and for the $W_L$ and 
$Z_L$ on the TeV brane we need to replace them with delta functions. We present 
resulting couplings in Table \ref{table} along with partial decay widths for dominant 
decay channels for the lightest KK (n=1) graviton which will be the focus of our analysis; 
see also \cite{Agashe:2007zd}. The $W_L W_L,Z_L Z_L$ and hh decay channels illustrate 
equivalence theorem once again, which is valid up to $(m_{W,Z}/m_G)^2$ where $m_G$ is the graviton mass.

Let us briefly explain the result for the  $\Gamma(G_n \to W_{KK} W_L)$ decay mode as 
it involves off-diagonal elements of the energy-momentum tensor of the gauge fields. 
Tha gauge boson mass matrix is \cite{Agashe:2006iy}:
\begin{eqnarray}
\frac{m_W^2}{2}\sum_{m,n=0}{a_{mn} A_{\mu}^{(m)}A^{\mu(n)}},
\label{mass}
\end{eqnarray}
and for the TeV brane Higgs scenario the off-diagonal elements $a_{0m}$ that describe the 
mixing of the zero and the mth KK mode $a_{0m}=\sqrt{2\pi kR}$. Thus, the off-diagonal 
elements of the energy-momentum tensor are given by:
\begin{eqnarray}
T_{\mu\nu}^{W(m,0)}&=&\eta_{\mu\nu}\left\{\frac{1}{2}F^{\rho\sigma(m)}F_{\rho\sigma}^{(0)}- m_W^2\sqrt{2\pi k R}A^{\rho(m)}A_{\rho}^{(0)} \right\}-\\ 
&&\left\{F^{\rho(m)}_{\mu}F_{\nu\rho}^{(0)}+F^{\rho(0)}_{\mu}F_{\nu\rho}^{(m)} - m_W^2\sqrt{2\pi k R}(A^{(m)}_{\mu}A_{\nu}^{(0)}+A^{(0)}_{\mu}A_{\nu}^{(m)})\right\}.  \nonumber
\label{EMT}
\end{eqnarray}

To the leading order in the $m_W/m_{KK}$, this results in the partial decay rate:
\begin{equation}
\Gamma(G_n \to W_{KK} W_L)=\frac{(C_{W_LW_{KK}G})^2 m_{G_n}^3}{480\pi}\cdot(1-\frac{m_{KK}^2}{m_{G_n}^2})\cdot\frac{m_W^2}{m_{KK}^2} \cdot f(\frac{m_{KK}}{m_{G_n}}),
\label{ZKKZ}
\end{equation}
where $f(x)\equiv r^2+(6r^2+20r+6)x^2+14(2-r^2)x^4+(6r^2-20r+6)x^6+r^2x^8$, $r\equiv \sqrt{2\pi k R}\approx 8.4$, and we neglected W boson mass in the phase-space consideration.

In the class of models we are working with, $m_1^G\approx 1.5 m_1^{KK}$ for the mass of 
the lightest KK graviton and the gauge fields \cite{Davoudiasl:2000wi} which translates 
into $f(2/3)\approx 173$. As graviton mass changes from 1.5 to 3 TeV (which will be the 
typical range for the graviton mass we consider in this paper), for our numerical estimates 
we take the $Br(G_1(2.25TeV) \to W_{KK} W_L)\approx Br(G_1(2.25TeV) \to Z_{KK} Z_L)\approx 
1/2 \times Br(G_n \to Z_L Z_L)$.

\begin{table}
\caption{Couplings of the first level KK graviton to the SM fields. The $t_R$ is assumed 
to be localized on the TeV brane. Parameter $m_1^G$ is the mass of n=1 graviton, $x_1^G=3.83$ 
is the first root of the first order Bessel function and $\epsilon\equiv e^{k\pi R}$. 
$N_c=3$ is number of QCD colors.}
\label{table}
\begin{center}
\begin{tabular}{|c|c|c|}
\hline
SM fields&$C_{qm1}$& Partial decay widths for n=1 graviton\\
\hline
gg(gluons)&$\frac{\epsilon}{2\pi kR}$& negligible\\
\hline
$W_L W_L$&$\epsilon$&$(cx_1^G)^2 m_1^G/480\pi$\\
\hline
$Z_L Z_L$&$\epsilon$&$(cx_1^G)^2 m_1^G/960\pi$\\
\hline
$t_R \bar{t}_R$&$\epsilon$&$N_c(cx_1^G)^2 m_1^G/320\pi$\\
\hline
h h&$\epsilon$&$(cx_1^G)^2 m_1^G/960\pi$\\
\hline
$W^{KK}W_L$&$\epsilon$&$390(cx_1^G)^2 m_1^G/960\pi \cdot (m_W/m_1^G)^2$\\
\hline
$Z^{KK}Z_L$&$\epsilon$&$390(cx_1^G)^2 m_1^G/960\pi \cdot (m_Z/m_1^G)^2$\\
\hline
\end{tabular}
\end{center}
\end{table}

The suppression in the coupling of the graviton to the gluons follows because the 
gauge boson has a flat wavefunction, and thus its couplings to the graviton is 
suppressed by the volume of the bulk $\pi kR\approx 35$. For the same reason, the 
decay of gravitons to transverse W and Z bosons, as well as photons, are suppressed 
by this volume factor. The masses of the KK gravitons are given by $m_n=x_n \mu$ 
where $x_n$ is n'th zero of the first order Bessel function.
Notice that we do not need $q\bar{q}$G coupling as it is Yukawa-suppressed, 
and graviton production is dominated by gluon fusion.

In this model the total width of the graviton is found to be 
$\Gamma_G=\frac{14(cx_1^G)^2 m_1^G}{960\pi}$ which is split between 6 dominant decay 
modes to $W_L W_L,Z_L Z_L,t_R \bar{t}_R$, hh, $Z_{KK} Z_L$, and $W_{KK} W_L$ in the 
ratio 2:1:9:1:0.5:0.5. Taking $c\sim 1$, 
the total graviton width is $\sim 7 \%\ $ of its mass and is very close to the 
corresponding width for RS KK Z$^{\prime}$ in the same model \cite{Agashe:2007ki}.

\section{Graviton spin measurement}

Now we discuss the strategy to confirm the spin-2 nature of the first KK mode of 
the RS graviton in our channels. Out of five possible polarization states of the 
graviton gluons can produce only $|J J_Z>=|2 \pm2>$ and $|J J_Z>=|2 0>$  states 
due to two facts: gluons do not have longitudinal polarizations and the total 
angular momentum has to be equal to J=2 (where we have chosen beam axis to be in 
the z-direction). Now, suppose that the two gauge bosons from graviton decay are 
produced at the polar angle $\theta$. We rotate the gluons-produced graviton state 
specified by polarization tensor $\epsilon_{\mu\nu}(J J_Z)$ by this angle \cite{Chung:1971ri}:

\begin{equation}
\epsilon_{\mu\nu}(2J_Z)=\sum_{J_Z^{\prime}} D^{(J)*}_{J_Z J_Z^{\prime}}(0,\theta,0) \epsilon^{\prime}_{\mu\nu}(2J_Z^{\prime}),
\label{rotation}
\end{equation}
where $\epsilon_{\mu\nu}(JJ_Z^{\prime})$ is the graviton state with the z-axis along 
the direction of the decay products, and 
$D^{(J)}_{J_Z J_Z^{\prime}}(\alpha,\theta,\gamma)\equiv <JJ_Z^{\prime}|R(\alpha,\theta,\gamma)|JJ_Z>=e^{-iJ_Z^{\prime}\alpha}d^{(J)}_{J_Z J_Z^{\prime}}(\theta)e^{-iJ_Z\gamma}$ is the Wigner D-matrix. 
Independent Wigner small d-matrix elements for the spin-2 state are 
presented in Appendix \cite{Berman:1965gi}. Now we may easily derive the angular dependence 
of the helicity amplitudes for our channels. They follow from Eq.\ref{rotation} for the $|2 \pm2>$ 
graviton state which is produced by $|+->$ and $|-+>$ gluons states:

\begin{equation}
\epsilon_{\mu\nu}(2\pm2)= d^{(2)}_{\pm 2 0}(\theta)\epsilon^{\prime}_{\mu\nu}(20)+ d^{(2)}_{\pm 2 1}(\theta)\epsilon^{\prime}_{\mu\nu}(21)+ d^{(2)}_{\pm 2 -1}(\theta)\epsilon^{\prime}_{\mu\nu}(2-1).
\label{decomp}
\end{equation}

Now just use Clebsch-Gordan decomposition of the $\epsilon^{\prime}_{\mu\nu}(20)$ 
and $\epsilon^{\prime}_{\mu\nu}(2\pm1)$ states in terms of 1$\otimes$1 final 
spin states to observe that, for example, helicity amplitude 
$A[g(\lambda_1)g(\lambda_2)\to Z(\lambda_3))Z(\lambda_4)]\equiv 
A_{\lambda_1 \lambda_2 \lambda_3 \lambda_4}$ for  
$A_{+-00}\sim d^{(2)}_{2 0}(cos\theta)$, $A_{+-0-}\sim d^{(2)}_{2 1}(cos\theta)$, 
and $A_{+-0+}\sim d^{(2)}_{2 -1}(cos\theta)$, where we have used Z boson in the final 
state for concreteness. Notice that we have not included 
the $d^{(2)}_{\pm 2 2}(\theta)\epsilon^{\prime}_{\mu\nu}(22)$ 
and $d^{(2)}_{\pm 2 -2}(\theta)\epsilon^{\prime}_{\mu\nu}(2-2)$ terms in Eq.\ref{decomp} 
as W and Z from graviton decay have longitudinal polarization and, thus, these terms cannot contribute.

Why $\epsilon_{\mu\nu}(2 0)$ graviton state does not contribute?
This is again due to the fact that a gluon is massless. If you allow a 
gluon to have mass, you will obtain additional helicity amplitudes proportional 
to the mass of the gluon in agreement with the above angular analysis. For example, 
someone would find that $A_{++00}\sim d^{(2)}_{0 0}(cos\theta)$, 
$A_{++0+}\sim d^{(2)}_{0 -1}(cos\theta)$, etc.  

Inherent to our analysis is the assumption that a graviton is produced essentially 
at rest so that its decay products are mostly back to back. The requirement to 
find the graviton center of mass frame will limit possible decay channels for 
the gauge bosons as we will see later. If the rest frame cannot be reconstructed,
we need to look for Lorentz invariant angular correlations which would encode
information on  the spin of the intermediate resonance. 
We do not pursue this approach here. 

Now we use the fact that the Wigner D-matrix elements $D^j_{mk}(\alpha,\beta,\gamma)$ 
form a complete set of orthogonal functions of the Euler angles $\alpha,\beta,\gamma$
(we use symbols $j$ and $J$ for the total angular momentum quantum number interchangeably):

\begin{equation}
\int_0^{2\pi} d\alpha \int_0^\pi \sin \beta d\beta \int_0^{2\pi} d\gamma \,\, D^{j'}_{m'k'}(\alpha,\beta,\gamma)^\ast D^j_{mk}(\alpha,\beta,\gamma) = \frac{8\pi^2}{2j+1} \delta_{m'm}\delta_{k'k}\delta_{j'j}, 
\label{orthogonal}
\end{equation}
to determine the spin of the resonance state.

Taking into account SM background events, we observe that measured normalized angular 
distribution is related to the graviton spin in the following way:

\begin{equation}
\frac{d\sigma}{\sigma dcos\theta}=\frac{\sum_{i=0,\pm 1} C_i \times [d^{(2)}_{2i}(cos\theta)]^2+\sigma_{backgd}(cos\theta)}{\frac{2}{2j+1}\sum_{i=0,\pm 1}{C_i}+\sigma_{backgd}},
\label{main}
\end{equation}
where we used the normalization for the Wigner small d-matrix, $C_i$'s are 
parton level cross section expansion coefficients convoluted over gluon PDF's, 
and we sum over three polarization states of the final state gauge bosons. 
As we will see for all the channels considered below, signal Z boson angular 
distribution peaks at pseudorapidity $\eta=0$ and, consequently, 
we will apply Eq.\ref{main} at this point.

\section{Applications}

We estimated SM background with the aid of the COMPHEP package \cite{comphepre}.  
For our graviton signal we used Mathematica program and partially cross-checked 
them with COMPHEP. CTEQ5M PDF's were used throughout (in their Mathematica distribution 
package \cite{Pumplin:2002vw} as well as intrinsically called by COMPHEP). 

%Proton level cross-section were obtained by convolving parton level cross section (which will be derived separately for every decay chann%el below) with gluon PDF's:
%
%\begin{equation}
%\sigma= \int dx_1 dx_2 f_g(x_1,Q^2) f_g(x_2,Q^2)\sigma(x_1 x_2 s).
%\end{equation}

\subsection{ZZ decay channel}

To see the method at work, let us start with the simplest example of 
$gg\to G_n \to Z_L Z_L \to e^+ e^-$, $\mu^+ \mu^-$ discussed in 
\cite{Agashe:2007zd} where the distinctive 4-lepton signal allows the 
reconstruction of all the masses of the particles in the decay chain. 
The dominant SM background for this purely leptonic mode is the $pp\to ZZ + X$, 
and the clean four-charged-lepton signal makes this mode 
a ``golden'' one.
%the absence of RS $Z^{\prime}$ coupling to the ZZ state  
For this process we have:

\begin{eqnarray}
A_{+-00}&=&A_{-+00}=\frac{s^2(\beta^2-2)sin^2\theta}{2}=\sqrt{\frac{2}{3}}s^2(\beta^2-2)d^{(2)}_{\pm 20}(cos\theta) \nonumber\\
A_{++00}&=&A_{--00}=0,
\end{eqnarray}
where $\beta=\sqrt{1-4M_Z^2/s}$ is the Z boson velocity.

\begin{figure}[htb]
\centering
\includegraphics[width=2.9in]{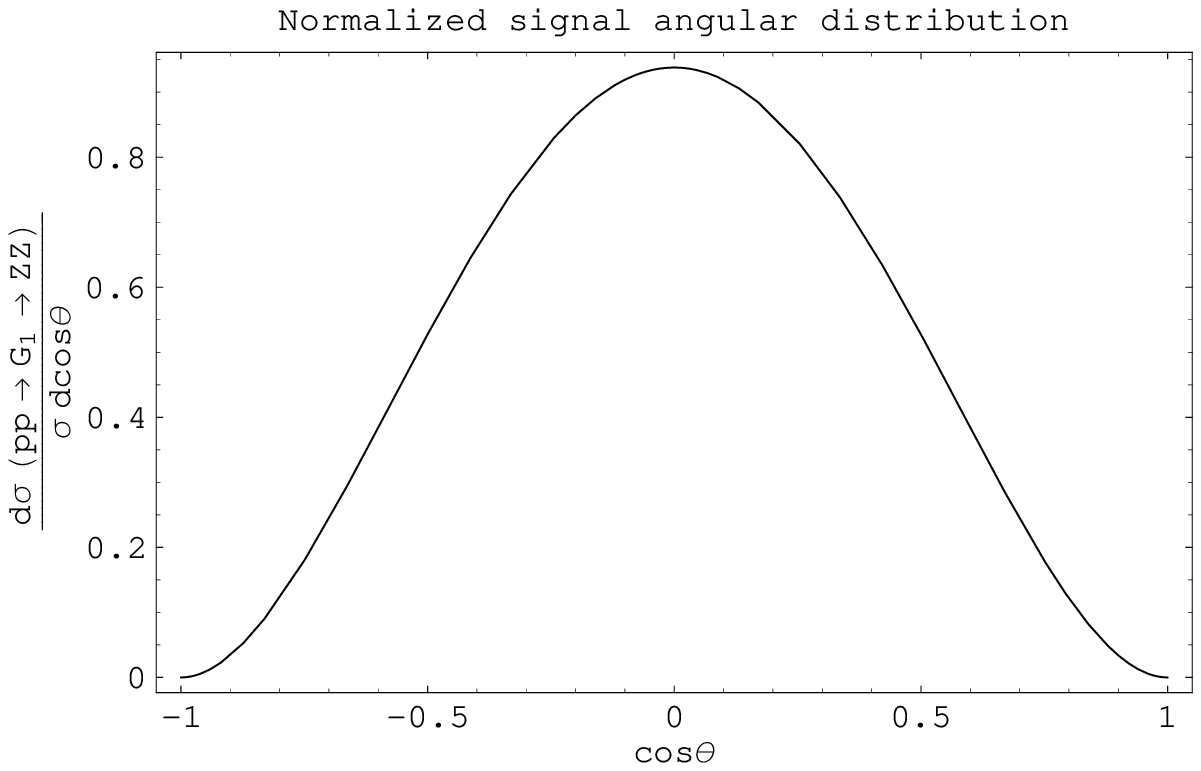} \includegraphics[width=2.9in]{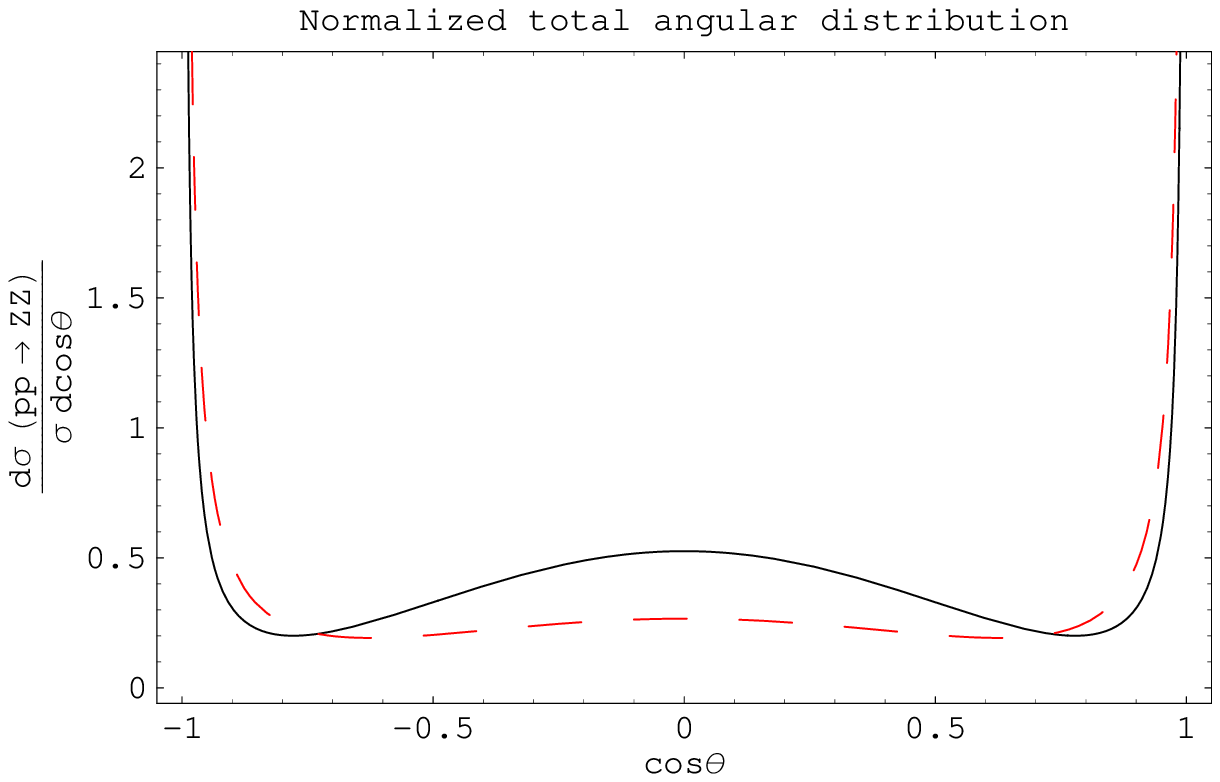}\\
\hspace{8mm}(a)\hspace{70mm}(b)
\caption{(a) Normalized angular distribution for the signal $\sigma(pp\to G_1 \to ZZ)$ 
cross-section due to the 1st KK graviton mode  and (b) Normalized total (i.e. including SM background) 
angular distribution for the $\sigma(pp\to  ZZ)$ cross-section for $m_1^G=1.5$ TeV (solid) 
and $m_1^G=3$ TeV (dashed) integrated in the $m_1^G\pm \Gamma_G$ ZZ invariant 
mass window with c$\equiv k/M_{Pl}=1$.}
\label{ppZZexample}
\end{figure}

Neglecting the SM background for a moment and using Eq.\ref{main}, we find that:

\begin{equation}
\frac{d\sigma_{signal}}{\sigma_{signal} dcos\theta}=\frac{2j+1}{2}d^{(2)}_{\pm 20}(cos\theta)d^{(2)*}_{\pm 20}(cos\theta)=\frac{5}{2}d^{(2)}_{\pm 20}(cos\theta)d^{(2)*}_{\pm 20}(cos\theta)\\,
\end{equation}
and, thus, the height of the peak in the normalized signal angular distribution 
as in Fig.\ref{ppZZexample}a is characteristic of the spin of the resonance. 
For our case, distribution peaks at $5/2 \times(\sqrt{6}/4)^2$=15/16 and is 
independent of the graviton mass.

The irreducible SM background to the$ ZZ$ final state is dominated by $q \bar{q}$ 
annihilation as gluon fusion proceeds via loop and, thus, interference of the KK 
graviton signal with SM background is negligible. The background cross-section 
exhibits forward/backward peaking due to t/u channel exchange while KK signal 
concentrates in the central rapidity region\cite{Agashe:2007zd}.

On Fig.\ref{ppZZexample}b we show the total (signal plus background) cross-section 
integrated in the $m_1^G\pm \Gamma_G$ $ZZ$ invariant mass window for two samples, 1.5 TeV 
and 3 TeV, graviton masses. For these mass values, we find SM background at cos$\theta$=0 
as $\sim 6 \%\ $and $\sim$23$\%\ $ of the signal, respectively. We observe that for 1.5 TeV 
case peak value at cos$\theta$=0 changed to $\approx$ 0.5 due to the fact that for this 
graviton mass $\sigma_{backgd}\approx \sigma_{signal}$ in $m_1^G\pm \Gamma_G$ $ZZ$ invariant 
mass window \cite{Agashe:2007zd} and, thus peak value reduced by $\sim$ half after 
normalization (see also Table \ref{tableZZ}). For the 3 TeV mass, the peak value $\sim$0.2 
as $\sigma_{backgd}\approx 3 \times\sigma_{signal}$. We may also impose pseudorapidity $\eta$ 
cut to reduce background, keeping the signal (almost) unchanged. For example, for the $\eta<2$ 
cut considered in \cite{Agashe:2007zd}, we find peak values as $\sim$0.8 and $\sim$0.5 for the 
1.5 TeV and 3 TeV graviton masses respectively. Also, the background may be further reduced 
using lepton angular distribution to distinguish longitudinally polarized $Z$ bosons from RS 
graviton decay from SM background \cite{Park:2001vk}.

Finally, using numerical results from Ref.\cite{Agashe:2007zd} and including $Z\to \tau^+ \tau^-$ 
channel not considered there, we obtain statistics presented in Table.\ref{tableZZ}. We assume 
100$\%\ $ efficiency for our clean 4-lepton signal. 
Poisson statistics CL to observe at least one signal event will be appropriate description if 
the number of background events $\lsim$ 10. We see that for 1.5 TeV and 3 TeV gravitons with 
1000 fb$^{-1}$ of data, we expect to have $\sim$ 130 events and $\sim$ 1 event respectively. 
This implies that higher luminosities are needed to reach 3 TeV graviton KK mass (for 3 ab$^{-1}$ 
SLHC discussed in the community, see for example \cite{Gianotti:2002xx}). The reason for optimism 
on the issue of the detection of the $\tau$'s from $Z$ decay is that $\sim$500 GeV energy $\tau$'s 
have a decay length of $l=\gamma \tau c\approx 20$ mm and therefore might leave visible tracks 
in the detector \cite{Bengtsson:1985ym}.

\begin{table}
\caption{Signal $pp \to ZZ \to$ 4 leptons cross-section (in fb) for the $m_G=1.5$ TeV 
and $m_G=3$ TeV with the corresponding leading SM background. Numbers in brackets 
correspond to $\eta<2$ cut case.
For the low number of events, $\lsim$ 10, Poisson statistics
is an appropriate description and the corresponding confidence level is, therefore, used. 
We assume 100$\%\ $ efficiency for our clean 4-lepton signal.  }
\label{tableZZ}
\begin{center}
\begin{tabular}{|c|c|c|c|c|c|}
\hline
1.5 TeV&No cuts&$\eta<2$ cut&$\#$ of events/1000 fb$^{-1}$&S/B&S/$\sqrt{B}$\\
\hline
Signal G $\to$ ZZ $\to$ 4 lept.  &0.13&0.13&130&1.3(6.5)&13(29)\\
\hline
SM ZZ$\to$ 4 lept. &0.1&0.02&100(20)&&\\
\hline
\hline

\hline
3 TeV&No cuts&$\eta<2$ cut&$\#$ of events/1000 fb$^{-1}$&S/B&CL\\
\hline
Signal G $\to$ ZZ $\to$ 4 lept.&0.001&0.001&1&0.33(1.25)&57$\%\ $(54$\%\ $)\\
\hline
SM ZZ$\to$ 4 lept. &0.003&0.0008&3(0.8)&&\\
\hline
\end{tabular}
\end{center}
\end{table}

\subsection{$W^{KK}(Z^{KK}) W_L(Z_L)$ decay channels}

For our next examples we need to consider the matrix element for the 
gg$\to G_n \to W^{KK}(Z^{KK}) W_L(Z_L)$ in the helicity basis. 
Working in the parton center of mass frame, the result is:
\begin{equation}
M(g^a g^b \to W_L W^{KK})= \frac{c^2}{2\pi kR \mu^2} \cdot  \frac{\sum_{\lambda_{1,2,3,4}}A_{\lambda_1 \lambda_2 \lambda_3 \lambda_4}\delta_{ab}}{s-(m_n^G)^2+i\Gamma_n^G m_n^G}
\end{equation}
where helicity amplitudes relevant for our process are: 

\begin{eqnarray}
A_{+-00}&=&A_{-+00}=\frac{\{\sqrt{2\pi k R}((r^2-1)^2 m_W^4-r^4 s^2)-4r^2 s m_W^2\}sin^2\theta}{2r^3}\label{hel}\\
%A_{+-+0}&=&A_{-++0}=\frac{2\sqrt{2s}m_W \{(r^2-1)^2 m_W^2-r^2 (r^2+1) s \}cos^3\frac{\theta}{2} sin\frac{\theta}{2}}{r^3}\nonumber\\
%A_{+--0}&=&A_{-+-0}=\frac{2\sqrt{2s}m_W \{(r^2-1)^2 m_W^2-r^2 (r^2+1) s \}sin^3\frac{\theta}{2} cos\frac{\theta}{2}}{r^3}\nonumber\\
A_{+-0+}&=&A_{-+0+}=\sqrt{\frac{s}{2}}\frac{m_W}{r^2}\left\{(\sqrt{2\pi k R}-1)m_W^2+(\sqrt{2\pi k R}+1)r^2 s\right\}(1-cos\theta)sin\theta\nonumber\\
A_{+-0-}&=&A_{-+0-}=\sqrt{\frac{s}{2}}\frac{m_W}{r^2}\left\{(\sqrt{2\pi k R}-1)m_W^2+(\sqrt{2\pi k R}+1)r^2 s\right\}(1+cos\theta)sin\theta\nonumber \\
A_{++00}&=&A_{--00}=A_{++0+}=A_{++0-}=A_{--0+}=A_{--0-}=0 \nonumber
\end{eqnarray} 
with $r\equiv m_W/m_W^{KK}$. To obtain $M(g^a g^b \to Z_L Z^{KK})$ replace $m_W$ with $m_Z$ . 
We have checked our results with Ref.\cite{Park:2001vk} where the process  gg$\to G_n \to Z Z$ 
was considered (which translates into r=1 in our notation), and we confirmed them. 

%Angular dependence of the amplitude follows from angular momentum analysis. Of the five polarizations states of the graviton (G), only th%ree states can be produced by gluons with $L_z=\pm 2,0$. $A_{+-00}$ and $A_{-+00}$ amplitudes correspond to $L_Z=\pm 2$  states and, thus%, $sin^2\theta$ is just manifestation of spherical harmonic $Y^{-2}_2\sim sin^2 \theta$. Sin$^2\hat{\theta}$ behavior of the amplitude al%so implies that our signal events will be concentrated in the central rapidity region. Also we notice that amplitudes are proportional to% masses of the bosons and, thus, they  vanish if at least one of final state bosons is massless because massless particle cannot have lon%gitudinal polarization. 

After straightforward calculation we arrive at the parton level cross-section:

\begin{equation}
\frac{d\sigma(gg\to W_L W^{KK})}{dcos\theta}= \frac{|M|^2}{512\pi s}(1-\frac{(m_W^{KK})^2}{s}),
\end{equation}
where we have neglected the W boson mass in the phase-space consideration.

As $W^{KK}$ and $Z^{KK}$ subsequently decay, we need to know their main decay 
channels which we now turn our attention to.

From now on, we generically call nth KK states of W and Z as $W^{\prime}$ and $Z^{\prime}$. 
We consider the simplified single bulk $SU(2)_L$ case and take $(t,b)_L$ to have close to a 
flat profile and $t_R$ on the TeV brane as they together do the best in satisfying the combined 
flavor-changing neutral currents (FCNC) and precision constraints.
After that, the decay widths for the leading channels of $Z^{\prime}$ 
and  $W^{\prime}$ are \cite{Agashe:2007ki} :

\begin{eqnarray}
\Gamma(W^{\prime} \to tb)&=&\frac{g_{SM}^2 m_{W'}}{16\pi}, \hspace{5mm} \Gamma(Z^{\prime} \to t\bar{t})=\frac{g_{SM}^2 (\kappa^2_V+\kappa^2_A) m_{Z'}}{4\pi c_w^2} \nonumber\\ \label{decay}
\Gamma(W^{\prime} \to W_L H)&=&\frac{g_{SM}^2 \kappa^2 m_{W'}}{192\pi}, \hspace{5mm} \Gamma(Z^{\prime} \to Z_L H)=\frac{g_{SM}^2 \kappa^2 m_{Z'}}{192\pi c_w^2}\\ 
\Gamma(W^{\prime} \to W_L Z_L)&=&\frac{g_{SM}^2 c_w^2\kappa^2 m_{W'}}{192\pi}, \hspace{5mm} \Gamma(Z^{\prime} \to W_L W_L)=\frac{g_{SM}^2 c_w^2\kappa^2 m_{Z'}}{192\pi}\nonumber,
\end{eqnarray}
where in the TeV brane Higgs scenario 
$\kappa\equiv g_n/g_{SM}\approx \sqrt{2\pi k R}\approx 8.4$ is the coupling strength 
of nth KK state relative to SM $SU(2)_L$ coupling, and $c_w$($s_w$) is the cosine (sine) of the Weinberg 
mixing angle. Notice that Ref.\cite{Agashe:2007ki} assumed Higgs 
as $A_5$ \cite{Contino:2003ve,Agashe:2004rs} 
and, thus, the IR brane coupling enhancement is equal to $\sqrt{\pi k R}$ there. 
Also, using the values of t$\bar{t}Z^{\prime}$ overlap integrals for the fermion 
profiles specified above \cite{Agashe:2007ki}, we obtain $\kappa_V\approx 1/4-5 s_w^2/3 $  and 
$\kappa_A\approx -1/4-s_w^2$. The enhancement of SM coupling for the decay channels in the last 
two rows of Eq.\ref{decay} follows from the fact that all the participating fields have a 
profile peaked near TeV brane compared to transverse zero-modes of W and Z, both having a flat 
profile in extra-dimension. Moreover, $W^{\prime} \to tb$ decay channel is not enhanced since  
$(t,b)_L$ fields have a close to a flat profile.
Corresponding branching fractions implied by Eq.\ref{decay} are presented in Table \ref{table1}.

\begin{table}
\caption{ Branching ratios of $W^{\prime}$ and $Z^{\prime}$ in the TeV brane Higgs scenario.}
\label{table1}
\begin{center}
\begin{tabular}{|c|c|c|}
\hline
Decay modes&$W^{\prime}$&$Z^{\prime}$\\
\hline
$W_L H$&0.51&-\\
$W_L W_L$&-&0.35\\
$W_L Z_L$&0.40&-\\
$Z_L H$&-&0.60\\
$t \bar{t}$&-&0.05\\
tb&0.09&-\\
\hline
\end{tabular}
\end{center}
\end{table}

\begin{figure}[htb]
\includegraphics[width=5.0in]{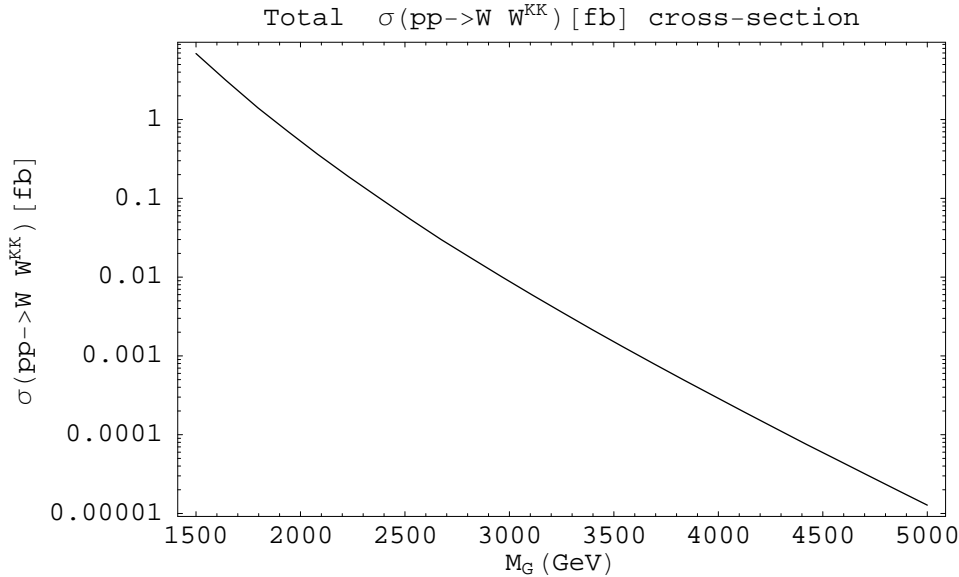}
\centering
\caption{(a) Total signal cross-section $\sigma(pp \to W_L W^{KK})\approx\sigma(pp \to Z_L Z^{KK})$ 
as a function of the 1st KK graviton mode mass integrated in the $m_1^G\pm \Gamma^G$ $W_L W^{KK}$ 
mass window and with c=1. }
\label{nocuts}
\end{figure}

We focus only on the 1st KK mode of W and Z as the effects of heavier KK modes are suppressed. 
On Fig.\ref{nocuts} we present the total resonant cross-section 
$\sigma(pp \to W_L W^{KK})\approx\sigma(pp \to Z_L Z^{KK})$ integrated in the 
$m_1^G\pm \Gamma^G$ $W_L W^{KK}$ mass window. Using branching ratios in Table \ref{table1} 
total cross-sections after KK state decays may be easily obtained.

Thus, the five possible final states are: $W_L W_L Z_L$ (which may come from both  
$Z_L Z^{\prime}$ and $W_L W^{\prime}$ intermediate states), tb$W_L$, $t\bar{t} Z_L$, $W_L W_L H$, 
and $Z_L Z_L H$. In this paper we will concentrate on $W_L W_L Z_L$ and $Z_L Z_L H$ states. $W_L W_L H$ 
final state faces the challenge to reconstruct efficiently the W mass from the W decay products. 
We will not consider tb$W_L$ and $t\bar{t} Z_L$ final states as both of them may additionally 
be produced through an s-channel KK gluon exchange which couples strongly to the $t\bar{t}$ pair.\\

{\bf \underline{ZZH decay channel}}\\

We are now in a position to discuss the more complicated case of the 
$pp\to G\to Z Z^{KK}\to ZZH$ final state where we have three independent helicity 
amplitudes involved (see Eq.\ref{hel}) compared to the above ZZ case where only one 
independent helicity amplitude survived. We assume that both Z's decay leptonically so 
that both Z masses can be reconstructed. Then, as one of the Z bosons comes directly 
from the graviton decay, it will have a bigger energy than the other one. We again 
would like to know how angular distribution of this Z may help to determine the spin 
of resonance its emitted from. We consider the ideal situation of pure signal events 
first and then add background events (which will depend on the mass of the Higgs) later.

\begin{figure}[htb]
\centering
\includegraphics[width=2.9in]{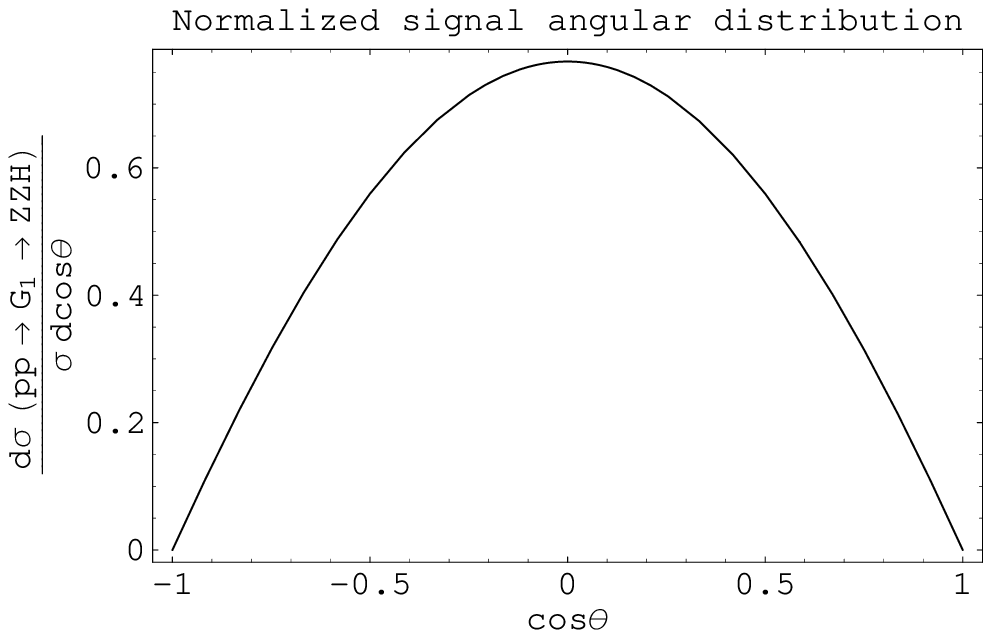} \includegraphics[width=2.9in]{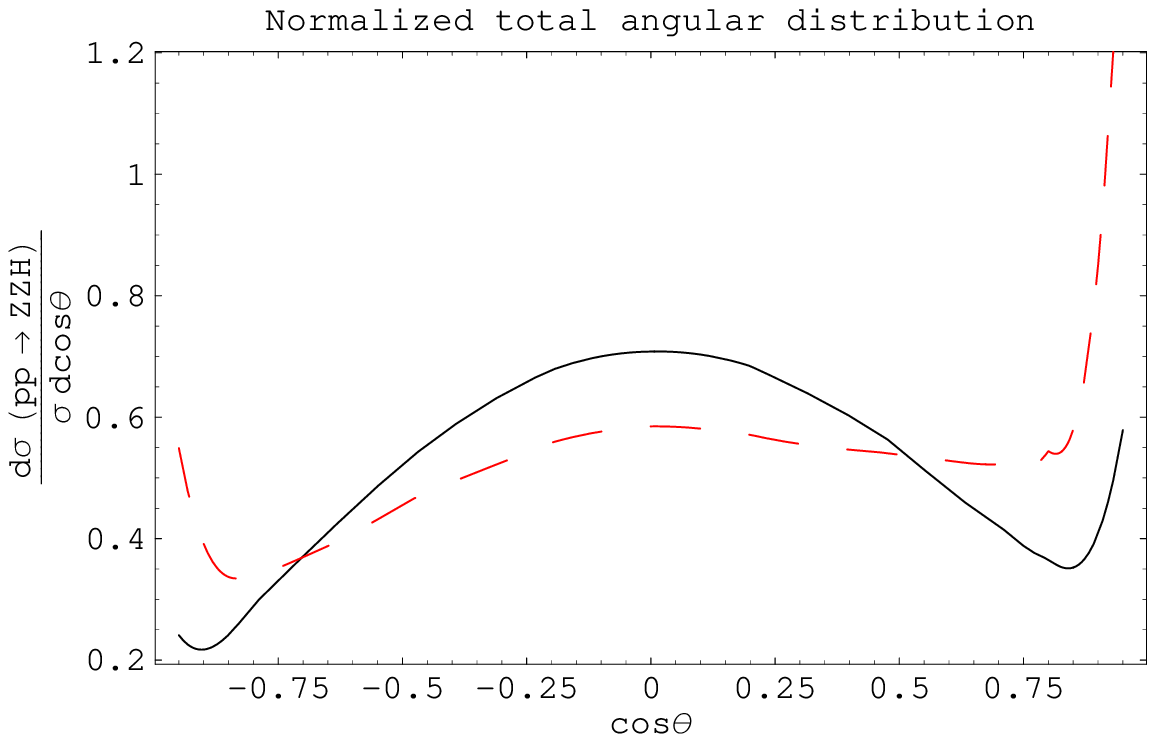}\\
\hspace{11mm}(a)\hspace{69mm}   (b)
\caption{(a) Normalized angular distribution for the signal $\sigma(pp\to G_1 \to ZZH)$ 
cross-section due to the 1st KK graviton mode and 
(b) Normalized total angular distribution for $m_1^G=1.5$ TeV (solid) 
and $m_1^G=2$ TeV (dashed) with cuts as in Eq.\ref{cuts}.}
\label{ppZZHexample}
\end{figure}

Again writing 
$d\sigma_{signal}/dcos\theta\equiv \sum_{i}C_i \times [d^{(2)}_{2i}(cos\theta)]^2$ and using $C_1=C_{-1}$, 
we have:

\begin{equation}
\frac{d\sigma_{signal}}{\sigma_{signal} dcos\theta}=\frac{C_0\times [d^{(2)}_{20}(cos\theta)]^2+C_1 \times([d^{(2)}_{2-1}(cos\theta)]^2+[d^{(2)}_{21}(cos\theta)]^2)}{\frac{2}{2j+1}\times (C_0+2C_1)},
\label{peak1}
\end{equation}
and, thus, the peak value occurs in this case at 
$\frac{(3/8C_0+C_1/2)}{\frac{2}{5}(C_0+2C_1)}\approx 0.77$. The normalized angular 
distribution is shown on Fig.\ref{ppZZHexample}a and again is independent of the 
mass of the graviton. Obviously, Fig.\ref{ppZZHexample}a applies to WWH and WWZ cases 
as well because the helicity amplitudes are the same. 

%In the former channel though we face the challenge to reconstruct efficiently the W mass while in the latter we have to keep in mind that% WWZ state can emerge from both $W^{KK}$ and $Z^{KK}$ decays.

Let us now consider the SM background for the ZZH case. It will depend on the leading 
decay mode(s) of the Higgs which, in turn, will depend on the mass of the Higgs boson. 
Important features can be highlighted by considering $m_H=120$ GeV case for which the 
leading Higgs decay mode is $H\to b\bar{b}$. Due to the large Lorentz boost of the Higgs, 
we expect 2 b-jets to merge; and, thus, we conservatively require to have 
4-leptons + 1 tagged b in the final state. 

\begin{table}
\caption{$pp \to ZZH \to$ 4 leptons $b \bar{b}$ cross-section (in fb) for the signal 
with $m_G=1.5$ TeV and $m_G=2$ TeV and the corresponding leading SM backgrounds with cuts 
as in Eq.\ref{cuts} and efficiency/rejection factors as discussed in the text. For the 
low number of events, $\lsim$ 10, Poisson statistics is an appropriate description and 
the corresponding confidence level is, therefore, used.}
\label{table3}
\begin{center}
\begin{tabular}{|c|c|c|c|c|}
\hline
1.5 TeV&Cuts and b-tag&$\#$ of events/500 fb$^{-1}$&S/B&CL\\
\hline
Signal G $\to$ ZZH $\to$ 4 leptons $+b\bar{b}$&0.0196&9.82&6.7&99.9$ \%\ $\\
\hline
SM ZZb$\to$ 4 leptons + b&1.6 $\times 10^{-4}$&0.08&&\\
\hline
SM ZZ$q_\ell \to$ 4 leptons+ $q_\ell$&1.5$\times 10^{-3}$&0.75&&\\
\hline
SM ZZg$\to$ 4 leptons + g&1.2$\times 10^{-3}$&0.6&&\\
\hline
SM ZZc$\to$ 4 leptons + c&5.6$\times 10^{-5}$&0.028&&\\
\hline
\hline

\hline
2 TeV&Cuts and b-tag&$\#$ of events/1000 fb$^{-1}$&S/B&CL\\
\hline
Signal G $\to$ ZZH $\to$ 4 leptons $+b\bar{b}$ &1.82$\times 10^{-3}$&1.82&1.36&61$ \%\ $\\
\hline
SM ZZb$\to$ 4 leptons + b&6.65$\times 10^{-5}$&6.65$\times 10^{-2}$&&\\
\hline
SM ZZ$q_\ell \to$ 4 leptons+ $q_\ell$&5.24$\times 10^{-4}$&0.52&&\\
\hline
SM ZZg$\to$ 4 leptons + g&7.28$\times 10^{-4}$&0.73&&\\
\hline
SM ZZc$\to$ 4 leptons + c&2.41$\times 10^{-5}$&2.41$\times 10^{-2}$&&\\
\hline
\end{tabular}
\end{center}
\end{table}

We consider 1.5 TeV and 2 TeV sample graviton masses and impose the following cuts:
\begin{eqnarray}
m_G=1.5 \hspace{2mm}TeV:&&  |\eta_{Z,H}|< 2, \hspace{2mm} m_{G}-\Gamma_G < M_{ZZH} < m_G+\Gamma_G \nonumber\\
m_G=2 \hspace{2mm}TeV:&&    |\eta_{Z,H}|< 2, \hspace{2mm} m_{G}-2\Gamma_G < M_{ZZH} < m_G+2\Gamma_G ,
\label{cuts}
\end{eqnarray}
where for 2 TeV case we doubled the ZZH invariant mass window to increase the number of events.

We use a b-tagging efficiency of 0.4 with a rejection factor for light jets ($u,d,s,g$) 
R=20 \cite{marchtalk}. We use a charm rejection factor $R_c=5$. In addition, we used 
BR($H\to b\bar{b}$)=0.7 and $\sum_{e,\mu,\tau}$BR$(Z\to \ell^+ \ell^-)\approx$ 0.1. 
All this results in the cross-sections presented in the second column of Table.\ref{table3}. 
We find a clear signal above the background for 1.5 TeV case and 80$ \%\ $CL for 2 TeV case. 
Notice that we used the  efficiency/rejection parameters optimized for low 
transverse momentum of the b-quark $P_{Tb}$, and 
rejection is expected to improve for high $P_{Tb}$ which is the case at hand. Also, 
efficient reconstruction of $Z$ mass from hadronic $Z$ decay will increase the number 
of signal events as those modes have a bigger BR. 

On Fig.\ref{ppZZHexample}b. we show the normalized angular distributions for 1.5 TeV 
and 2 TeV graviton masses considered. We observe that for 1.5 TeV mass 0.77 peak value 
remains (almost) unchanged as the distribution is dominated by signal events, 
while for the 2 TeV mass value peak is less distinct.\\ 

{\bf \underline{WWZ decay channel}}\\

As discussed above, signal angular distribution is the same as in Fig.\ref{ppZZHexample}a 
for this case because the helicity amplitudes are the same. Additionally, $WZ$ or $WW$ 
invariant mass presumably should have resonant (due to $W^{\prime}$ or $Z^{\prime}$) 
distribution; but we don't impose cuts on $WZ$ or $WW$ mass as we would like to keep our 
analysis as general as possible. At this point we have to decide on the decay modes of 
$W$ and $Z$ boson. We again allow $Z$ decay leptonically so that we reconstruct $Z$ mass 
efficiently, and we use the angular distribution of this $Z$ for determination of the 
spin of the graviton. Now, if we allow both $W$'s decay hadronically, due to the huge 
Lorentz boosts of these $W$'s, we pick up 2 leptons + 2 jets as a background for our 
decay mode which we find to be overwhelmingly bigger than our signal. Thus, we  
use $(W\to jet)$ (W$\to$ leptons) and (Z$\to$ leptons) as our final state.

For the leptonic $W$ decay, due to small angular separation between missing neutrino 
and charged lepton, we may estimate longitudinal (L) component of the $\nu's$ momentum as: 

\begin{equation}
p_{\nu}^{L}\approx\frac{\slashed{E}_T p_l^{L}}{p_{T_l}}.
\end{equation}
Using this collinear approximation, the momentum of the leptonic W is reconstructed and, thus, 
 we can calculate the (presumably) resonant invariant mass of the WW or WZ system. 
In doing so, we assumed that leptons are coming from the W decay as the reconstructed leptonic W mass will be zero in the collinear approximation.
Also notice that in this approximation, the $M_{WW}$ measurement error for the TeV energy W bosons is $\sim m_W/E_W\sim$ 0.1.

We again consider 1.5 TeV and 2 TeV sample graviton masses and impose the following cuts:
\begin{eqnarray}
|\eta_{Z,W}|< 1, \hspace{2mm} m_{G}-\Gamma_G < M_{WWZ} < m_G+\Gamma_G.
\label{cuts1}
\end{eqnarray}

We have to remember that Eq.\ref{main} is valid only if we integrate over whole angular 
coverage of the detector. Fortunately, as our signal concentrates in central rapidity 
region, even such a hard pseudorapidity cut changed the signal cross-section for both 
graviton masses only by about 8$\%\ $which is in the range of experimental uncertainties.
This cut also changed the peak value in the normalized angular distribution of 
Fig.\ref{ppZZHexample}a from 0.77 to 0.83 value.

In addition, we use the result of \cite{Agashe:2007ki} which finds that jet mass cut:
\begin{equation}
65 \hspace{2mm}GeV <M_{jet}< 115 \hspace{2mm}GeV
\label{cuts2}
\end{equation}
achieves acceptance fraction of 0.78 for the signal and 0.3 for the background events. 
Table.\ref{table23} shows our results after all this cuts are imposed.

\begin{table}
\caption{$pp \to WWZ \to$ 3 leptons + jet + $\slashed{E}_T$ cross-section (in fb) for 
the signal with $m_G=1.5$ TeV and $m_G=2$ TeV and the corresponding leading SM 
backgrounds with cuts as in Eq.\ref{cuts1} and Eq.\ref{cuts2} and 
efficiency/rejection factors as discussed in the text.}
\label{table23}
\begin{center}
\begin{tabular}{|c|c|c|c|c|}
\hline
1.5 TeV&Cuts&$\#$ of events/300 fb$^{-1}$&S/B&S/$\sqrt{B}$\\
\hline
Signal G $\to$ WWZ $\to$ 3 leptons + jet + $\slashed{E}_T$&0.10&30&1.16&5.9\\
\hline
SM WWZ$\to$ 3 leptons + jet + $\slashed{E}_T$&0.0026&0.78&&\\
\hline
SM WZ$q \to$ 3 leptons + jet + $\slashed{E}_T$&0.0656&19.7&&\\
\hline
SM WZg$\to$ 3 leptons + jet + $\slashed{E}_T$&0.018&5.4&&\\

\hline
\hline

\hline
2 TeV&Cuts&$\#$ of events/1000 fb$^{-1}$&S/B&S/$\sqrt{B}$\\
\hline
Signal G $\to$ WWZ $\to$ 3 leptons + jet + $\slashed{E}_T$&0.008&8&0.26&1.44\\
\hline
SM WWZ$\to$ 3 leptons + jet + $\slashed{E}_T$&6.8$\times 10^{-4}$&0.68&&\\
\hline
SM WZ$q \to$ 3 leptons + jet + $\slashed{E}_T$&0.023&23&&\\
\hline
SM WZg$\to$ 3 leptons + jet + $\slashed{E}_T$&0.0072&7.2&&\\
\hline
\end{tabular}
\end{center}
\end{table}

\begin{figure}[htb]
\includegraphics[width=5.0in]{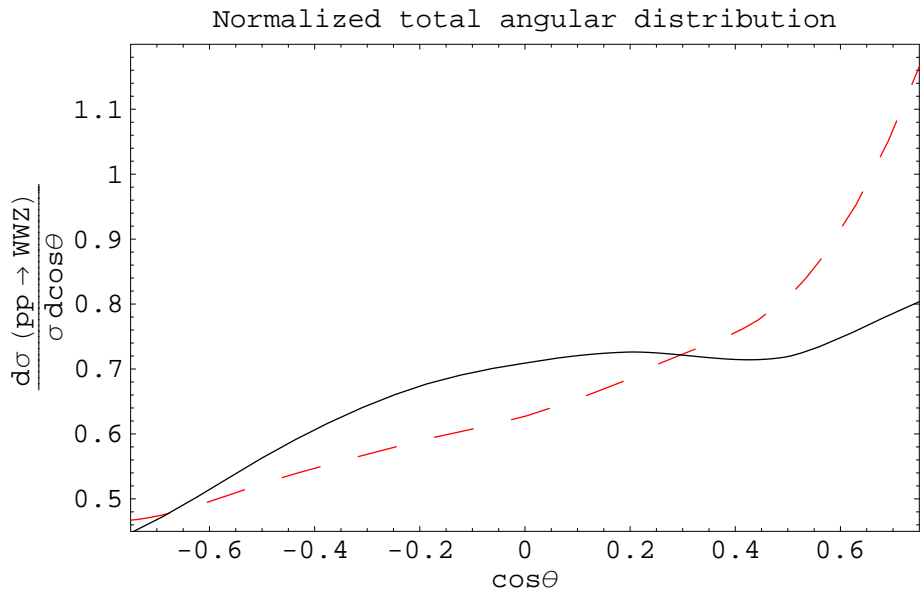}
\centering
\caption{ Normalized total angular distribution for $\sigma(pp\to WWZ)$ cross-section 
for $m_1^G=1.5$ TeV (solid) and $m_1^G=2$ TeV (dashed) with cuts as in Eq.\ref{cuts1} 
and Eq.\ref{cuts2}.}
\label{WWZ}
\end{figure}

Finally, on Fig.\ref{WWZ} we show the normalized angular distributions for the 1.5 TeV 
and the 2 TeV graviton masses considered. We observe that for the 1.5 TeV mass 
peak value of 0.83 for the case of zero background changed to about 0.7. 
This value can also be obtained applying Eq.\ref{main}
and using the fact that for this mass $\sigma_{backgd}\approx \sigma_{signal}$ 
as can be seen in Table.\ref{table23} and the fact that 
$\sigma_{signal}(cos\theta=0)\approx 1.68\sigma_{backgd}(cos\theta=0)$. For the 2 TeV mass 
value, peak is no longer seen due to the dominance of the background. 

\section{Conclusions}

In this work, we have extended earlier studies of the discovery potential of warped 
gravitons at the LHC which concentrated on the gravitons decaying into the ``gold-plated'' 
$Z_L Z_L$ channel, $W_L W_L$, channel and into the $t\bar{t}$ pair. We have considered 
resonant production of the first RS KK graviton mode via gluon-fusion process followed by 
its subsequent decay to $W^{KK}(Z^{KK}) W_L(Z_L)$ and $Z_L Z_L$ pairs. We focused on 
confirmation of the unique spin-2 nature of the graviton using Z boson angular distribution 
in the graviton rest frame for all these modes. We performed angular analysis using  Wigner 
D-matrix in order to derive the relationship between the graviton spin, angular distribution 
peak value, and other theoretically calculable quantities. As our method only requires to 
measure this peak value, where most of the signal events will be concentrated, it may be possible 
to achieve this goal with a relatively low sample of $O$(10) events. In any case, our main aim 
in this work is to illustrate how our method can work, at least in principle. Using statistical 
results of \cite{Agashe:2007zd} for $pp \to ZZ \to$ 4 leptons and our analysis of 
$pp \to W^{KK} W \to WWZ \to$ 3 leptons + jet + $\slashed{E}_T$ decay modes, we 
showed that with 1000 fb$^{-1}$ of data, these channels allow this number of events to accumulate 
for the RS graviton up to $\sim$ 2 TeV. As a byproduct of our analysis, we found that 
$W^{KK}(Z^{(KK)}) W_L(Z_L)$ graviton decay modes, which have not been studied before, have 
a Br comparable to the zero mode decay channels as summarized in Tables.\ref{table}-\ref{table23}. 
These decay modes open new channels to search for the RS gravitons. As an extra bonus, 
reconstruction of intermediate KK gauge bosons in these modes will be important 
to reveal the detailed workings of the RS model.

\begin{acknowledgments}

We thank Hooman Davoudiasl for a careful reading of the manuscript and for
many useful discussions. 
Work of O.A. is supported in part by DOE under contract number DE-FG02-01ER41155. A.S. 
is supported in part by the DOE grant DE-AC02-98CH10886 (BNL).

\end{acknowledgments}

\appendix

\section{Spin-2 Wigner small d-matrix}

\begin{eqnarray}
d^{(2)}_{22}(\beta)&=&\frac{(1+cos\beta)^2}{4},\hspace{5mm} d^{(2)}_{21}(\beta)=-\frac{1+cos\beta}{2}sin\beta, \hspace{5mm} d^{(2)}_{2-1}(\beta)=-\frac{1-cos\beta}{2}sin\beta\nonumber\\
d^{(2)}_{20}(\beta)&=&\frac{\sqrt{6}}{4}sin^2\beta,\hspace{10mm} d^{(2)}_{2-2}(\beta)=\frac{(1-cos\beta)^2}{4},\hspace{9mm} d^{(2)}_{10}(\beta)=-\sqrt{\frac{3}{2}}sin\beta \hspace{1mm}cos\beta \nonumber\\
d^{(2)}_{11}(\beta)&=&\frac{1+cos\beta}{2}(2cos\beta-1),\hspace{10mm} d^{(2)}_{1-1}(\beta)=\frac{1-cos\beta}{2}(2cos\beta+1),\nonumber\\
d^{(2)}_{00}(\beta)&=&\frac{3cos^2\beta-1}{2}.
\label{Wigner}
\end{eqnarray}

\end{document}